# Are Conversational AI Agents the Way Out? Co-Designing Reader-Oriented News Experiences with Immigrants and Journalists

Yongle Zhang
College of Information, University of Maryland
College Park, Maryland, USA
yongle@umd.edu

Ge Gao
College of Information, University of Maryland
College Park, Maryland, USA
gegaoumd.edu

## Abstract

Recent discussions at the intersection of journalism, HCI, and human-centered computing ask how technologies can help create reader-oriented news experiences. The current paper takes up this initiative by focusing on immigrant readers, a group who reports significant difficulties engaging with mainstream news yet has received limited attention in prior research. We report findings from our co-design research with eleven immigrant readers living in the United States and seven journalists working in the same region, aiming to enhance the news experience of the former. Data collected from all participants revealed an "unaddressed-or-unaccountable" paradox that challenges value alignment across immigrant readers and journalists. This paradox points to four metaphors regarding how conversational AI agents can be designed to assist news reading. Each metaphor requires conversational AI, journalists, and immigrant readers to coordinate their shared responsibilities in a distinct manner. These findings provide insights into reader-oriented news experiences with AI in the loop.

## 1 Introduction

Mainstream news disseminated in a given society constitutes an essential venue for people living there to stay informed and make life plans. Among them, a subgroup who frequently experience struggles with news reading are immigrants.

Previous research has documented various barriers that limit immigrants' information seeking in their societies of living. Commonly noted ones include lack of full fluency in the majority language [2, 22], insufficient understanding of the local cultural environment [3, 21, 36], and mismatches between common knowledge shared among local majorities and immigrants' own lived situations [69, 85]. Recent work has found that, to compensate for their lack of engagement with mainstream news, immigrants often consume ethnic media (e.g., [19, 105]) or turn to networks with other immigrants (e.g., [46, 74]). However, such workarounds can raise high-stakes concerns, as these alternative information may be unsustainable, vary in quality, and engender significant echo chambers.

In response to a recent call for creating reader-oriented news experiences [86], a handful of HCI scholars have discussed the design of news technologies to serve immigrant readers. For example, Gao et al. proposed redesigning automated translation systems to better support immigrants' news consumption [36]. Similarly, Heuer and Glassman developed an accessible text framework that guides the restructuring of news content for immigrants, as well as for other populations with reading difficulties [42]. Zhang et al. studied how immigrants seek comprehension support from LLMs while reading news articles [107]. They concluded that technology should enable these individuals to critically assess news content rather than merely offering language support. Collectively, these projects have contributed valuable knowledge for enhancing immigrants' news experiences through design solutions.

An inherent problem of this body of work, though, is that immigrant readers are often treated in isolation. They are recognized as facing challenges in news reading [3, 36, 107], and are then provided with technical tools to cope with the challenges in an entirely self-directed manner. Other stakeholders in the news ecosystem — particularly journalists, whose work strongly influences the production of mainstream news content for its readership [10, 11, 39, 65] — have rarely been considered.

The current research aims to extend the empirical understanding as well as the design thinking for immigrants' news experiences by connecting their voices with those of journalists. Our investigation involved 11 immigrant readers and 7 journalists from the same metropolitan area in the U.S. capital region. They participated in a co-design study that we organized, which included a series of interlinked activities spanning multiple days. Such a process enabled everyone at the study sessions, including ourselves, to recognize the relationship between values prioritized by immigrant readers and those emphasized by journalists. All participants also collaboratively envisioned technology designs for immigrant readers by situating their design speculations within existing links between news production and consumption.

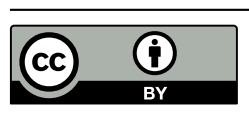

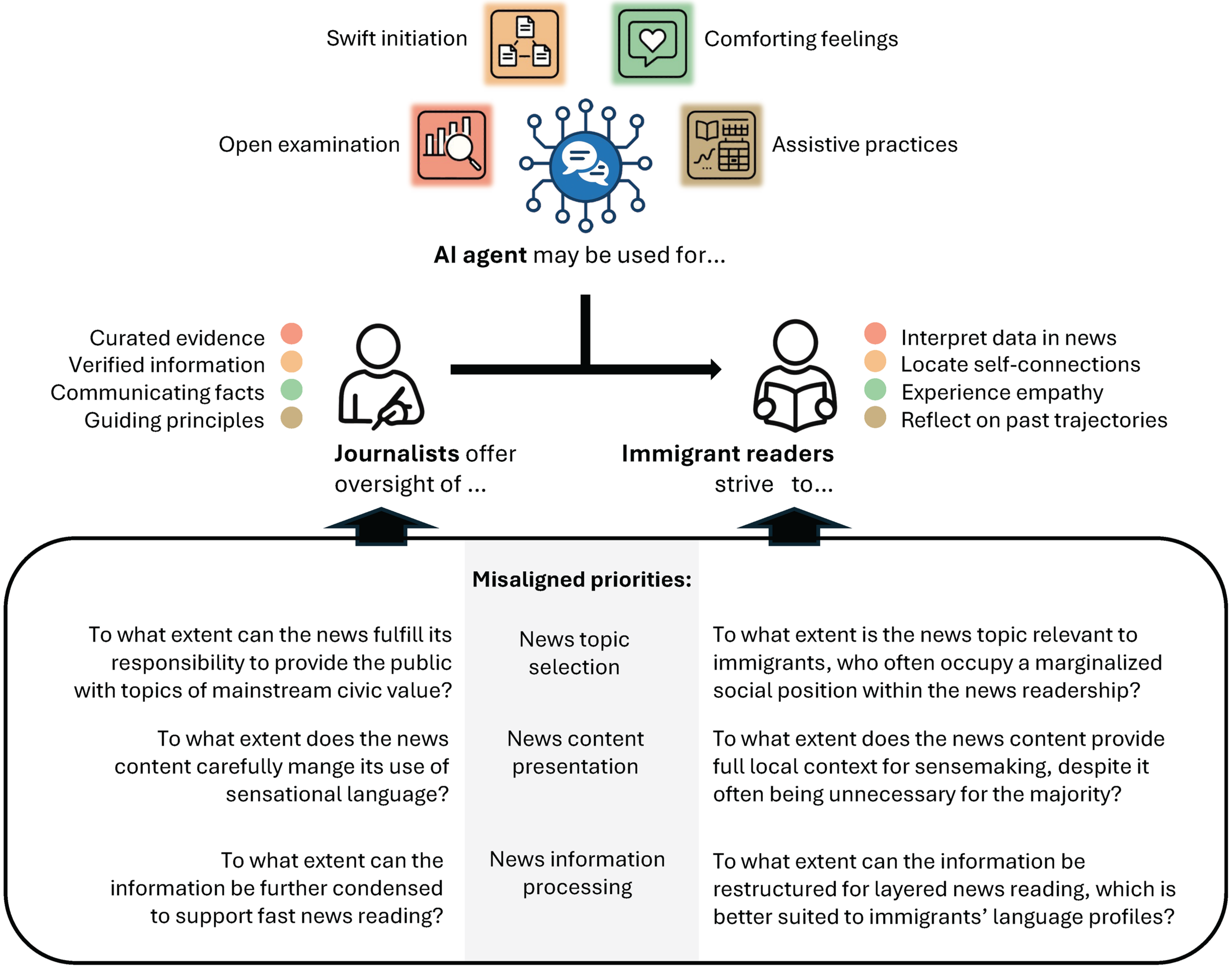

**Figure 1: A visual summary of all primary findings of our current research. The top section illustrates four participant-proposed design metaphors connecting journalists, readers, and conversational AI agents for enhancing immigrants' news experience. These metaphors consider the AI agent as Data Decoder, Connection Informer, Empathetic Friend, and Trajectory Witness, respectively. The lower section illustrates value misalignments that are inherent across journalists and immigrant readers, which contextualize the human-AI coordination workflow implied by each design metaphor.**

Our findings surfaced multiple sets of (mis)alignments between immigrant readers and journalists on values that should be prioritized in the selection of news topics, the presentation of news content, and the processing of news information. Grounded in these insights, participants articulated four metaphors for future news technologies serving immigrant readers: as a data decoder, as a connection informer, as an empathetic friend, and as a trajectory witness. These metaphors all frame conversational AI agents as the default technology for use. Across metaphors, the AI agent never functioned alone; rather, it coordinated with journalists in four distinct ways to balance effectiveness and accountability (see Fig.1). This research contributes meaningful insights for enhancing immigrant readers' news experience by appropriately aligning roles among technology, journalists, and immigrant readers themselves.

## 2 Related Work

We reviewed three bodies of literature that motivated our current work: computational tools for news consumption, news reading among immigrants, and news production by journalists. Syntheses of this prior knowledge led us to adopt co-design as our research device to explore aspirational connections between immigrant readers, journalists, and technology design toward reader-oriented news experiences. We conclude the section by presenting research questions that guide our investigation.

### 2.1 Computational Tools for News Reading

As news reading today shifts from traditional media to digital platforms, a variety of commercial applications have been developed to support this evolving practice. From online aggregators (e.g., Google News, Apple News) to automated fact-checkers (e.g., FactCheck.org,

NewsGuard), readers' accessing, navigation, and evaluation of news content have become increasingly mediated by technology and more effective.

In the research domain, recent advancements in natural language processing and machine learning have led to a surge of computational tools in the news space. Some of these tools focus on comprehension, such as automated summarization or guided reading assistance across given news content and other textual materials (e.g., [20, 25, 33]). Others aim to assist with error detection, enabling users to verify claims and evaluate information quality (e.g., [5, 29, 66, 73]). A third line of work addresses media bias by helping individuals recognize political leanings in news reports and their framing strategies (e.g., [67, 98, 103, 104]).

While such efforts point to a promising future for technology-powered news experiences, they have largely concentrated on well-bounded tasks with limited attention to a reader's organic news journey in everyday life. To this latter end, empirical studies of human behaviors have shown that modern news consumption is highly situation-dependent, shaped by diverse backgrounds of the readers, and spans various topics and formats ([12, 102]). Mismatches between technologies developed in labs and readers' real-world practices sometimes prompt concerns, as reported during user studies sessions of existing tools. For example, news readers have expressed worries about inherent algorithmic biases of the tools [66, 103], difficulties stemming from steep learning curves [98], and gaps between a tool's specific use cases and people's habitual news reading practices [25, 98]. These issues underscore the need to bridge news technology design with the people who stand to benefit from it through a more human-centered approach.

## 2.2 Key Stakeholders in the News Ecosystem

The news ecosystem involves individuals from various stakeholder groups. Two groups central to our research are immigrant readers and journalists. They occupy distinct yet interrelated positions in news consumption and production.

*2.2.1 News Consumption by Immigrant Readers.* Immigrants constitute a persistent yet often overlooked readership group for mainstream news in a given society. Prior studies have documented important differences in how immigrants and local majorities engage with news. One classical finding, repeatedly confirmed across multiple projects, considers the purposes driving people's mainstream news reading: immigrants often harness news information to meet settlement and survival needs, whereas locals' news consumption is more frequently linked to higher-end demands such as civic participation [4, 21, 69, 81].

More detailed contrasts have been reflected in research on news sensemaking and interpretation [3, 28, 43, 107]. For instance, Alencar and Deuze conducted interviews with immigrants who relocated to the Netherlands and Spain from Latin America, Sub-Saharan Africa, and Eastern and Southern Europe. They found that the interpretations these participants derived from mainstream news often deviated from those of locals [3]. Gao et al. conducted experience sampling studies with Chinese individuals living in the U.S. and Japan. Participants' information-seeking logs revealed a lack of depth, as well as blurred focus, in their sensemaking of COVID-related news in both countries [36]. Zhang et al. examined questions asked by Asian immigrants and locals in the U.S. when reading the same set of mainstream news articles. Their data analysis showed that immigrants, compared to locals, faced significantly greater challenges in connecting the given news content with other ongoing issues in society [107].

Many scholars attribute immigrants' struggles with news reading to their limited knowledge of the local environment and imperfect fluency in the majority language (e.g., [1, 36, 64, 85]). Several projects in the broader HCI and CSCW communities have also proposed systems to support immigrants' adoption of local knowledge, as well as language learning, in everyday contexts (e.g., [18, 76, 90, 110]). To date, however, little concrete guidance has been offered to design news technologies that address the specific needs of immigrant readers.

*2.2.2 News Production by Journalists.* Journalists oversee the production of news content that is ultimately consumed by the public. As detailed in numerous ethnographies and media sociology studies, this process is shaped by established professional norms and routinized practices [35, 61, 72, 77, 94].

In particular, Tuchman's fieldwork with journalists and editors working in the U.S. newsrooms documented these people's everyday routines and scheduled practices [92–94]. It demonstrates that news is not merely a neutral reflection of reality but the result of professional assessment and constructions of what is deemed newsworthy. Shoemaker and Reese synthesized empirical research on news production [75, 79, 80]. They proposed a multi-layer hierarchical framework, showing that news content is not determined by an individual journalist's or reporter's taste. Instead, it must be codified through professional routines, accessible resources, and factors operating at the ideological level. Kovach and Rosenstiel analyzed interviews with journalists and media leaders [56]. They identified a set of principles guiding journalism practices, including truth-seeking and independence.

Over time, news production has gradually moved toward a reader-oriented view, although without being entirely defined by it [6, 78, 86]. Modern practice in this domain considers local citizens as active participants who may contribute, disseminate, and provide feedback on information that goes into news [16, 30, 82, 87]. Through readers' forum columns on newspapers [95], call-in radio programs [8], call-out surveys [41], and online comments attached to digital articles [15, 89], journalists have become more attentive than before to readers' voices. — Yet, from the standpoint of our current research, these existing journalist-reader exchanges rarely involve immigrants.

## 2.3 Reader-Oriented News Experience via Co-Design

With the above sets of literature in mind, we argue that human-centered technology design for enhancing immigrant readers' news experiences should integrate insights from immigrants and journalists as a collective. Taking this approach does not mean deciding whose views are more important or legitimate than others. Rather, our goal is to recognize common ground across the two stakeholder groups and to experiment with bridging their gaps through collaborative efforts.

A small but growing body of recent HCI research has explored technology design for news experiences (e.g., [24, 45, 108]). However, synthesized perspectives across stakeholder groups received little coverage in the literature, despite their value. One fashion of this research has relied on experiments and usability interviews to test innovative forms of technology use in news reading. These studies are typically conducted in controlled settings, where participants, acting as news readers, try out system prototypes with specific features or mock-up interfaces [14, 24, 45, 107–109]. Another body of studies has been more design-driven and exploratory. They often conducted design activities with stakeholder representatives to outline conceptual designs for future news technology (e.g., [53, 68, 101]). Notably, the stakeholders participating in these design processes were largely limited to the news production side.

In this sense, our current work complements prior literature by engaging both immigrant readers and journalists in co-design explorations grounded in the former's needs. Following the tradition of research through design [34, 37, 112], this methodological choice enables us to generate knowledge about immigrant readers' news consumption practices, as well as to uncover organic connections between those practices and corresponding actions on the news production end [9, 37, 55, 111, 112]. Artifacts produced in this process, typically including narratives, scenarios, concepts, and sketches, are not only design outcomes; they also function as "boundary objects" that promote constructive dialogue across stakeholder groups, each representing a unique position in the problem domain under study [37, 58, 83].

We proposed two research questions (RQs) to guide our investigation through co-design with immigrant readers and journalists. Both RQs address essential insights for successful news technology design that neither group could generate independently:

***RQ1***: What is the relationship between the values prioritized in news consumption by immigrant readers and those emphasized in news production by journalists? Where do they align and/or misalign?

***RQ2***: How do immigrant readers and journalists envision the design of technology for reader-oriented news experiences of the former? What roles do they each play in relation to the technology?

Next, we present the setup of our co-design activities, which provides "a third space [62, 63]" that transcends traditional boundaries to enhance news experiences of immigrant readers.

# 3 Method

We conducted co-design with 11 immigrant readers and 7 journalists, following the design process model proposed by Fails et al. [32]. The entire research process consists of three successive phases (see Fig.2): onboarding sessions prior to the formal study, activity set one ("News on My Radar"), and activity set two ("News Without Barriers"). Activity set one and two each included two sub-activities: during the former sub-activity, participants were tasked with producing certain materials based on our requests; these materials then enabled us to better prepare for interactions at the latter sub-activity. Participants were compensated with $20 USD per hour via gift cards. They were also reimbursed for transportation expenses for any in-person parts of our activities. The entire research procedure was reviewed and approved by the University Institutional Review Board. We provide full details of the arrangement of our activities below.

## 3.1 Participants Recruitment and Composition

All participants were from the same metropolitan area in the U.S. capital region and were comfortable with English, enabling them to attend joint in-person activities for the best quality of communication during the research process. Among them, immigrant participants were recruited through two channels: a) printed flyers distributed at grocery stores, wellness centers, public libraries, and other public facilities within the region; and b) digital flyers posted to community platforms including Nextdoor, Craigslist, and relevant Reddit forums. Journalists were contacted through professional email lists and direct messages on LinkedIn. Table 1 provides demographic information about participants from the two stakeholder groups.

## 3.2 Co-Design Activities

*3.2.1 Onboarding sessions.* We arranged an onboarding session with every participant, either in person or over Zoom, depending on the participant's preference. Following practices in other co-design work (e.g., [40, 48, 109]), having this step prior to the formal research activities helped establish rapport with participants, and also supported their expectation setting and understanding of upcoming activities.

In these sessions, the lead researcher introduced our research team, provided an overview of the project, and then conducted the consent process with participants. Afterward, the researcher probed each immigrant reader to describe their daily news habits and technology usage; each journalist had a similar conversation with the researcher about their overall work routine in the news industry. The researcher took notes to document useful information to understand the person's background and prepare later activity plans. Each participant also completed a demographic survey before leaving. The onboarding session took about 1 hour on average.

*3.2.2 Activity Set One: "News On My Radar".* Through this set of activities, we aimed to explore the relationship between values prioritized by immigrant readers and those emphasized by journalists (RQ1). We arranged two sub-activities, drawing on prior co-design research that highlights the benefit of having participants bring personal objects from daily life to spark narrative exchanges [26, 27, 31, 76, 99].

For the first sub-activity, each immigrant reader was tasked to keep a one-week diary about their daily news practices. Specifically, we encouraged participants to take written notes or screenshots of the news they attended to in life. Every two days, participants received an email prompting them to reflect on why they were drawn to the documented news and how they made sense of it. Participants were not required to share their diaries daily.

For the second sub-activity, we paired immigrant readers and journalists into in-person small groups based on shared availability. Most small groups consisted of 1 journalist and 2 immigrant readers, and one group included 1 journalist and 1 immigrant reader. During these small group sessions, immigrant readers shared their news diaries after anonymization steps assisted by the researcher.

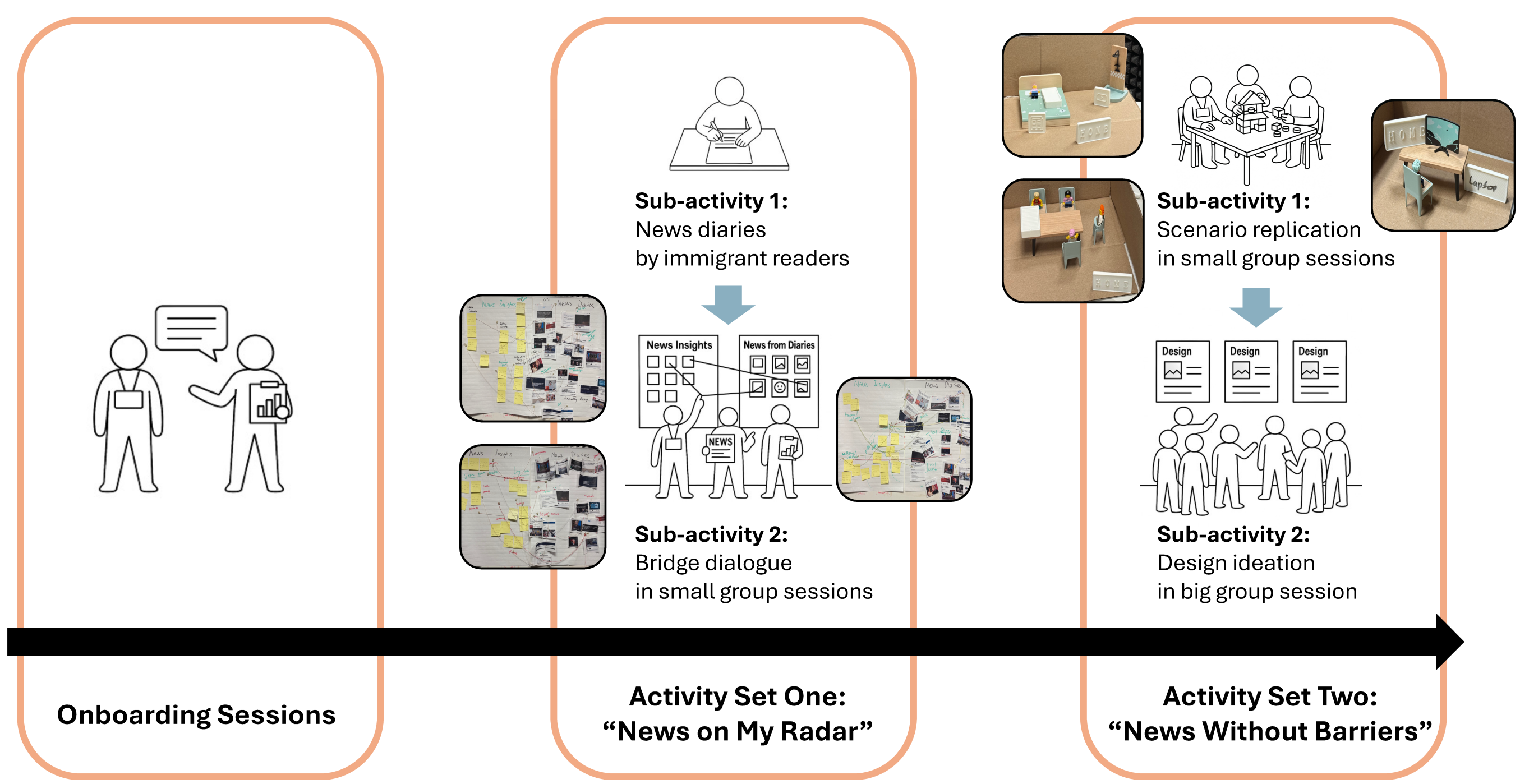

**Figure 2: An illustration of the entire research process consists of three successive phases: onboarding sessions, activity set one titled 'News On My Radar,' and activity set two titled 'News Without Barriers.' During activity set one, our immigrant participants kept a one-week diary of a self-selected news pieces they read in daily life; these news collections were used as design probes to anchor and elicit small-group dialogues between immigrants and journalists (see the three news collection boards for examples), allowing them to hold organic discussions and/or piece together sticky notes to exchange their news values. During activity set two, participants were provided with LEGO sets and other tangible materials to replicate where, how, and with whom their daily news reading occurs (see the three sets of LEGO crafts for examples), supporting the recovery of more situated experiences of immigrants' news reading; these scenarios, along with the materials generated in activity set one, were then displayed again in large-group discussions to help contextualize and scaffold the co-design of news technologies among all participants. To protect participants' self-initiated ideas, researchers did not restrict access to any materials they had generated throughout the research process. Participants could leverage anything they deemed useful to support any stage of their design thinking and conversations.**

**Table 1: Demographics of All Participants**

| Immigrant Readers | | | | | | Journalists | | | |
|---|---|---|---|---|---|---|---|---|---|
| **ID** | **Age** | **Gender** | **Country of Origin** | **Years of Immigration** | **Employment Status** | **ID** | **Age** | **Gender** | **Featured Experiences Besides Reporting** |
| **I-1** | 18 | Male | Ethiopia | <5 years | Student | **J-1** | 24 | Female | Virtual set producer |
| **I-2** | 18 | Male | Ethiopia | <5 years | Student | **J-2** | 34 | Female | Digital media specialist |
| **I-3** | 18 | Female | Ethiopia | <5 years | Student | **J-3*** | 28 | Female | Media and audience analyst |
| **I-4** | 26 | Male | India | <5 years | Full-time | **J-4** | 24 | Female | Video editor |
| **I-5** | 28 | Female | India | <5 years | Student | **J-5** | 41 | Female | Senior producer |
| **I-6** | 26 | Male | India | <5 years | Student | **J-6** | 35 | Female | Digital media producer |
| **I-7** | 25 | Male | Vietnam | <5 years | Full-time | **J-7** | 37 | Female | Senior digital content producer |
| **I-8** | 27 | Male | Vietnam | <5 years | Full-time | | | | |
| **I-9** | 28 | Female | Hong Kong | 5–10 yrs | Full-time | | | | |
| **I-10** | 40 | Male | Mainland China | >15 years | Freelance | | | | |
| **I-11** | 61 | Female | Mainland China | <5 years | Retired | | | | |

*Note: J-3 participated only in the big group session under Activity Set Two.*

All participants in the same small group session then engage in bridge dialogue to distill insights from the sample news pieces and exchange thoughts. These insights could include, but were not limited to, commonalities and differences across news pieces, notable attributes of each news piece, and any elements that participants liked or disliked from reading the news piece. Each small group session lasted 1.5-2 hours. The researcher facilitated the entire activity. They audio-recorded the session with participants' permission.

*3.2.3 Activity Set Two: "News Without Barriers".* Through this set of activities, we engaged participants from both stakeholder groups in collectively envisioning future technology design that addresses immigrant readers' needs (RQ2). Again, we arranged two sub-activities on different days to best support our participants' transition into design ideation.

The first sub-activity occurred on the same day as each small group session under "News on My Radar." In particular, we provided participants with LEGO sets, craft materials, and 3D-printed objects to replicate their daily news reading scenarios. Such hands-on experiences have been proven effective for quickly immersing participants in lived scenarios of their past experience (e.g., [50, 57, 88, 100]) and, in the context of our research, revealing representative challenges they encountered in news reading. This sub-activity took about 30-40 minutes per small group. The researcher audio-recorded participants' scenario and challenge descriptions, and photographed those LEGO-constructed scenarios.

The second sub-activity happened one week later. During that transition week, the research team worked on preliminary analysis of data obtained from small group sessions. They also synthesized the materials — including narratives, sample news pieces, and photographs — that could elicit participants' design speculations at the upcoming big group session. The big group session involved all participants. In this session, they reviewed materials generated from the week before, shared their reactions on immigrant news reading challenges contextualized in those materials, discussed potential technological solutions, and detailed how their speculated technologies should function in relation to both immigrant readers and journalists. The big group session lasted about 6 hours in total. Multiple members of the research team joined to facilitate. They audio-recorded the session and collected design storyboards generated at the session for later analysis. No video recording was obtained, in line with participants' requests.

## 3.3 Data Analysis

We followed the guidance of the thematic analysis method to examine our data inductively [17]. Following an iterative process, the first and last authors of this paper independently conducted open coding on an initial subset of the data to generate preliminary codes. The two people then met to discuss these codes, resolve discrepancies, and apply shared understandings to the coding of the next batch of data. They continuously compared codes across individuals, groups, and activities to identify recurrent concepts and emerging relationships. A third external reviewer was invited to review our codes and the themes connecting them, which helped ensure the rigor and credibility of our analysis. This cycle of coding, comparison, and refinement continued until no new conceptual insights emerged. The following section presents findings synthesized from the co-design.

# 4 Findings

The co-design activities with our participants, followed by careful analysis of their data, yielded inspiring insights into enhancing reader-oriented news experiences. The remainder of this section presents our primary findings concerning two aspects. We begin by comparing the fundamental assumptions held by journalists and immigrant readers regarding what's valued in news production and what immigrants demand from news reading, respectively (RQ1). We then describe representative challenges highlighted by immigrant readers. For each challenge, we synthesize participants' visions of how conversational AI agents may help enhance the news experiences of immigrant readers, often through coordination with journalists in specific ways (RQ2). These two sections of findings were synthesized from our analyses of activity set one (Section 3.2.2) and activity set two (Section 3.2.3), respectively.

For clarity of presentation, we use italics within quotation marks to indicate exact excerpts from our participants' self-reports. We also assign corresponding icons to differentiate segments of findings by how they relate to each stakeholder group: marks alignments across journalists and immigrant readers; alone indicates what applies primarily to journalists but not to immigrant readers; and alone for vice versa.

## 4.1 News Matters But How: (Mis)alignments Between Journalists and Immigrant Readers

While all participants considered mainstream news an essential venue connecting people with the society they live in, the narratives they shared revealed clear alignments and misalignments between the two stakeholder groups. Through inductive analysis of their self-donated reflections and news materials, we identify three sets of (mis)alignments with respect to news topic selection, news content presentation, and news information processing. These findings were primarily derived from participants' exchanges during the "News on My Radar" activities. They provide essential knowledge that underpins our planning for, and insight extraction from, the "News Without Barrier" activities.

*4.1.1 Drivers of Topic Selection Among All Options.*

***Popularity attracts attention.*** We invited journalists and immigrant readers to join small group sessions, reviewing all the news pieces collected in the latter's one-week diary. This part of the activity naturally prompted participants in the same session to discuss why certain topics were frequently featured in news stories, while others were not. A shared view, repeatedly mentioned across all small groups, was that news is part of today's attention economy; it thrives on the public's interest.

Topics that represent *"trends in the pop culture here"* and/or *"cared by most people [in this society]"* were thus strongly valued in both the production and the consumption of news. For journalists, reports about famous public figures, sports events, or gasoline price changes in the U.S. are almost always good choices for writing up a news article because *"they attract more views, and more views mean more revenue, which is a must for news as a business."* Such reports also received substantial coverage in our immigrant participants' news diaries — not always because immigrants held a genuine interest in those public figures or events, but more for the purpose of social learning and adaptation. As many immigrant readers in our study explicitly pointed out, *"following what's at the center of attention of the local majority is very important [because] it prepares us to adapt to this society."*

***"Because it is news."*** Satisfying popularity alone will not explain all the topic selections made on the news production side. Another parallel value driving journalists' choices is to provide the public with *"what they should be aware of."* Following this value sometimes requires a journalist to argue strongly for their proposed topics to be approved by editors. For instance, groundwater contamination in the U.S. is considered an issue of civic value by many journalists and news outlets; however, most people may not want to read about it when other options are available. J-6 shared her thoughts on why advocating for these *"hard topics"* matters for journalism as a profession. This sentiment was echoed by other journalists among our participants as well:

> *"News is not all about pleasing the public, although that's apparently important. When we studied journalism, we were taught that news carries social responsibility. The saying goes, 'when there is no news, it is good news.' It means not everything we cover will be pleasant to read. Still, we covered it because it is news." [J-6]*

***"It is likely to affect immigrants."*** Our immigrant participants also possess other rules of thumb that guide their selection of news topics to read. The dominant one considers the extent to which a given topic relates to the immigrant group. To our surprise, we noticed a clear emphasis on civic rights in immigrant readers' reflections on their own news reading; this challenges the impression we previously formed during literature search, which primarily discussed immigrants' instrumental needs. Participants at our research sessions reported paying close attention to sustaining their *"opportunities for self-advocacy," "political well-being,"* and *"freedom of speech"* in this country. That said, because immigrants are not the target audience of most mainstream news, they often have to spend extra effort investigating the exact connections between a news piece and themselves. As explained by I-9:

> *"The political environments surrounding immigrants and citizens are different even though [we live] in the same country. We want to think twice about news updates on employment, healthcare, education, and so on. Some of these updates can mean different things to us, like how I can protect my rights [in those domains] or how I can contribute. Most news won't directly talk about this, and most readers won't care, but will it affect immigrants?" [I-9]*

#### 4.1.2 Strategies of Content Presentation for Sensemaking.

***Credibility comes before everything else.*** As participants reviewed news pieces documented in the immigrant participants' diaries, they occasionally asked each other questions such as *"Does this content make sense to you?"* and *"Why is this news written in this way?"* A second theme emerging from these bridge dialogues concerns what each stakeholder group values in news content presentation. Credibility appears to be a common, foundational requirement underscored by both journalists and immigrant readers.

This emphasis on being credible is closely connected with participants' daily exposure to society-wide discussions on *"misinformation"* and *"fake news."* For journalists, stating traceable facts to support claims and opinions has become today's default norm of reputable news practice. The rise of digital media equips them with convenient tools, especially in-text hyperlinks and embedded screenshots, to clarify essential information sources. The same mindset is adopted by news readers as well. When quoted information is linked back to the original source, immigrant readers interpreted it as *"a good indicator of high-quality content that I can trust."* Several participants also added that they sometimes manually check whether the same information and its sources were mentioned across multiple news outlets. To them, *"if the presentation of a news piece allows easy cross-verification, we build quick trust in the news content."*

***"Be careful with sensational expressions."*** Admitting the necessity of ensuring credibility, multiple journalists made this insightful comment that having credible content *"does not mean the news report is non-leading or neutral."* Rather, any content can be presented with sensational language; that is, expressions that are emotionally appealing and reaction-provoking. The effects of such a presentation strategy may go both ways. When used appropriately, sensational language enhances a journalist's ability to capture readers' attention on urgent matters. On the flipside, it can also be utilized as a technique to *"fabricate clickbait"* and to *"make credible or traceable information serve an idea the reporter is pushing."* J-7 elaborated on this latter point by referring to some news pieces in immigrant participants' diaries:

> *"Here, the title says, 'Will DOGE layoffs even make a dent?' or 'As tariffs cause chaos…'. The news article does provide data and cases with sources in its content, but the title indicates an attitude. It leads readers to interpret those data in certain ways. … In practice, I think no news can be 100% objective or bias-free. But we are trained to be very careful in choosing the way to present [the content], [and] to be careful with [being] sensational." [J-7]*

***"What's put in a rich context wins out."*** While journalists' caution against sensational presentation is understandable to immigrant readers, we did not hear them initiate much reflection on this exact point. In fact, a few immigrant participants explicitly stated that they *"are not sensitive to the connotation of those adjectives and expressions in English"* because it is in their second language. For these individuals, what makes their news reading effort payoff is not scrutinizing the tacit meaning behind certain language styles, but rather encountering the news in a rich context that facilitates sensemaking. This value is clearly reflected in the coverage of video-based news pieces among immigrant participants' diaries. Many were created by individual YouTubers or TikTokers. The video usually includes a grassroots *"reporter"* who describes recent news by connecting the immediate event to its broader social contexts, as well as by grounding the narratives in the person's own lived experiences. I-11 explained what he appreciated about such videos, in a way that mirrors the perspectives of other immigrant participants:

> *"I am following many YouTubers, some in English and some in other languages [I speak]. Part of the routine of their channels is to talk about local news featured in U.S. mainstream medias. Their content is a mixture of describing the news, usually with [explanations on] where the issue originates from or the local context I didn't really know, and adding their own comments on it. It's very easy for me to resonate this way, or to just stay cool about it . . . because I can easily tell how much this news overlaps with things I know." [I-11]*

*4.1.3 Levels of Information Processing With Constraints.*

***Optimization for constraints.*** Last but not least, conversations at all small group sessions continuously reminded all participants that news reading in today's information world *"is always constrained."* Such constraints may stem from a person's intermittent availability, limited cognitive resources outside daily work, lack of ubiquitous access to big-screen devices, or some combination of these. Participants from both stakeholder groups found it is unrealistic to expect people to *"do very in-depth processing"* or *"digest many details"* when engaging with news at most moments of the day, if at all. Journalists repeatedly mentioned that a major decision they face on the production side is how to *"optimize the amount of information going into a news piece."* This aligns with shared views among our immigrant participants, who noted that *"it's easy to just browse for a main idea, but any steps beyond that take up time."*

***"Take things out, or people won't read it."*** To adapt to those constraints, journalists often choose to include only a subset of the whole information they have gathered during behind-the-scenes investigations. Participants described the specific criteria determining what is filtered in or out as *"highly dependent."* Nevertheless, some common heuristics were reported by more than one person, such as the one J-2 pointed out:

> *"I always do a lot of research — checking several databases, interviewing different people, [and] collecting all sorts of perspectives — when drafting my [news] article. But not all that will be included [in the published version]. From the very first journalism class, we were told to 'always think about your audience.' Especially, think of the audience like they are tired engineers who come back home after a long day, sit on the couch, and open a newspaper, or turn on the TV, or get on social media. You do not want to give them an extra hard time." [J-2]*

***"One piece of information at a time."*** Somewhat in deviation from journalists' assumptions, our immigrant participants all believed that they are able to manage complex information processing despite limitations in availability and such. The solution, specifically, is to distribute their mental load by staging it across steps. A frequently mentioned reading flow begins with *"scrolling through a list of news during breakfast or before bed for main ideas,"* followed by *"more serious reading or extended searching on what's important when I'm more free."* All immigrant participants also reported that they had tried using AI to assist this reading flow, similar to how they leveraged AI to process bulky documents at work. As I-5 reported:

> *"Sometimes I see news that I'm curious to know more about but don't have time [for deep reading] immediately. I just copy and paste the entire thing into GPT or similar tools and get a summary. It gives me a quick sense of whether I should remember to read [anything in this news] more carefully later. . . When I want to do extended reading [beyond the given news content], I use GPT to get recommendations too." [I-5]*

***In sum***, our findings above suggest that journalists and immigrant readers do not always align in their views regarding what is valued in the production versus the consumption of mainstream news. While journalists strived for careful gatekeeping grounded in their professional training, immigrants place greater emphasis on situating news reading within their everyday lives — lives that often feature their unique social statuses in the society, varying levels of familiarity with the local culture, and distinctive rhythms or paces of reading in the majority language.

## 4.2 Co-Designed Metaphors for Conversational AI Agents in Challenging News Scenarios

The bridge dialogues in the "News on My Radar" activities consolidate the ground for both stakeholder groups to delve into the most salient issues raised by our immigrant participants, and to envision solutions collectively. Our data analysis, primarily based on the "News Without Barrier" activities, identified four representative challenges described by immigrant readers. For each challenge, participants envisioned a specific form of coordination connecting technology, immigrant readers, and journalists. Across these exchanges, technology consistently emerged as conversational AI agents, which, in our view, reflects the pervasive influence of ChatGPT and similar tools in contemporary everyday life. We structured our findings around the four metaphors of agent design synthesized through the analysis. While these design metaphors are presented in a linear order, participants did not rank any as more pressing than the others.

*4.2.1 The Data Decoder.* Given their often-limited fluency in the society's local cultures as well as the majority language, immigrant readers identified numerical data cited in news reports as their best entry point for making sense of the news. They described such data as *"more accessible"* and *"more objective"* than verbal descriptions. Many further commented that reading numbers exempts them from the comprehension challenges posed by local references, which they have frequently encountered in mainstream news.

However, just like verbal expressions can be written in sensational ways, numerical data may also be framed to contain meanings beyond the numbers themselves. Our immigrant participants were acutely aware of this latter point. One challenge discussed extensively at the big group sessions was how to enhance readers' decoding of the data cited in news reports. We detail the thoughts exchanged between stakeholder groups in the context of a news piece that they frequently referred to for grounding their discussion. The news piece stated, *"Survey shows 40% of D.C. restaurants are* *likely to close by the end of the year,"* and it used this data to support the claim that the D.C. area was experiencing severe financial distress due to rising supply costs and unemployment. Immigrant readers envisioned that a conversational AI agent could act as a data decoder, assisting them in delving into the source dataset from which this "40%" originated. Journalists also contributed to this discussion, with particular reflections on how the speculated reader-agent interaction may mirror their own investigative work with data during news production.

***Immigrant-agent conversations to interpret data in news.*** News articles today, especially those published by reputable outlets, often include hyperlinks or notes specifying the source of the data they cite. But immigrant readers want more than just a link. The following excerpt of their technology ideation at big group sessions offers a closer look into what our immigrant participants hoped the AI agent could communicate with them:

> *"I-6: I definitely want to know more about where the source dataset is. AI can tell us more [about that dataset], but I want to make sure it does not just summarize. We don't want important details to be missing.*
> *I-4: Right. We don't want AI to just provide an overall image. For example, if there were 200 restaurants surveyed here and 40% of them are closing, that would mean 80 restaurants. I don't think D.C. only has 200 restaurants. It seems low. So, can AI tell me what defines 'a restaurant' here?*
> *I-8: Yeah, it's about how they sampled. We want a clearer picture. If there are actually 2,000 restaurants in total, that means something totally different [from what the current news means].*
> *I-4: Also, what type of restaurants are these? Maybe they serve a food that's no longer in demand. Food habits change. If the survey was done by a public association, they may provide more details [in the source dataset]. The AI can point us to that for a more complete image of the D.C.'s restaurant ecosystem."*

The central idea underneath this conversation, and in others like it, was that the AI agent should help immigrant readers better understand the nature of the source dataset. All immigrant participants agreed on three essential aspects in this regard: size of the data, composition, and sampling strategy. They wanted explanations not just about the source dataset referenced in the news article, but also about other comparable sources that the AI agent could identify. In addition, several immigrant participants underscored the importance of historical trends in interpreting numbers. As explained by I-1:

> *"It always requires context to know whether a number is big or small, like the 40% of restaurant closing, or the $25 billion investment said in another [news piece]. Are those a lot? Are there significant changes? It would be useful to know how the same thing was like in the past, if there are some records about it." [I-1]*

***Journalist for curated evidence, AI for open examination.*** The assistance that immigrant readers seek from talking to an AI agent reminds journalists of their own use of AI in curating news content, including the data being cited. Journalists noted that commercial tools, such as ChatGPT, are already part of their practice for identifying *"candidate datasets to study"* during news production. That said, they all strongly objected to the idea of relying on AI without personally reviewing the datasets.

For these journalists, human oversight must accompany every single number cited in news, since *"most news articles do not have the room for many numbers, and that makes those few numbers [going into the article] carry a lot of weight."* They explained why it could be risky to let an AI agent interpret numbers, and the source datasets behind them, entirely on behalf of people. J-2 and J-1, for instance, shared with immigrant participants the roots of their concerns about AI:

> *"AI can surely talk to people and answer their questions about the data. What worries me, though, is which perspectives will get highlighted in those responses and the hidden choices behind them. There are articles published by the Atlantic Council, talking about how thousands of propaganda messages were produced with malicious purposes. AI tools were using that information in their training and ended up spreading that propaganda. Most people won't be aware [that] this is happening. AI may have a lot of power over what people think." [J-2]*

> *"It makes sense to explore the data with AI. But you want to make sure that what's explored are facts, not explanations of 'why this reporter cited this data but not the other.' Reporters usually have looked into different datasets, and we sometimes have data teams working with us to choose the right data to include. We may leave out some data because it's sponsored by institutes with specific perspectives, or it's hard to verify under time pressure. There can be many valid reasons. But AI may speculate, 'this report is biased.' Then it confuses readers rather than clarifying the meanings. " [J-1]*

Through all these exchanges, immigrant readers and journalists converge on a future scenario, where a conversational AI agent serves as a **data decoder** in coordination with human stakeholders. In this arrangement, an immigrant reader initiates their conversation with the AI agent to request factual descriptions about a source dataset related to the news they read. The AI agent answers the reader's questions while also detailing steps the reader can take to verify each answer themselves. Part of the new report's metadata contains notes produced and authorized by the journalist. These notes do not appear in the published article, but the AI agent can access them to prepare for *"why-questions"* raised by the reader, such as *"What makes the referred data a more valid choice for this news piece?"* In this way, the journalist and the AI agent combine their strengths in explaining already-curated evidence and enabling more open examination, respectively, to help immigrant readers decode the meanings behind data in news.

*4.2.2 The Connection Informer.* With sufficient content comprehension, immigrant participants are eager to grasp the real-world implications of a news piece. In particular, they shared prior experience of encountering news that suggested incoming, high-stakes adjustments to their everyday lives, such as *"care pathway updates during COVID"* or *"job market shifts over the last year."* As individuals who must navigate a different set of regulations in healthcare, employment, education, and many other areas compared to local citizens, they constantly wonder, *"Am I taking the right actions in light of what the news just said?"*

There is, unfortunately, no easy answer to this question in most cases. Several immigrant participants described the difficulty with the notion of *"dominoes"*: the news signals that some changes will bring consequences for immigrants down the line, but *"nobody outlines a concrete connection to you until it's already late."* A substantial portion of the big group discussions focused on this challenge. In the same format as the previous section, we synthesize design ideas exchanged between stakeholder groups in a frequently cited context: U.S. mainstream news about taxes. Immigrant readers speculated that a conversational AI agent could serve as a connection informer, highlighting connections between tax-related administrative updates and actions they might need to take in life. Their thoughts prompted journalists' reflections on the inherent *"popularity-inclusivity"* tension within societies characterized by diverse populations.

***Immigrant-agent conversations to locate self-connections.*** The conversations our immigrant participants envisioned with an AI agent closely resemble existing interactions between users and ChatGPT. Key steps in this interaction included an immigrant reader providing the URL of a news piece, or URLs of multiple news pieces they considered related, and asking the AI agent to specify what the news meant for them in actionable terms. The difficulty, in their view, was not obtaining a response, but obtaining one that was adaptive to the reader's immigrant status and included calibrations based on their personal circumstances.

Most immigrant participants initially suggested setting up a personal profile when using the AI agent. This profile would contain basic demographic information, similar to a Facebook profile, allowing content to be curated for the person's interests and needs. However, debate soon arose about how personal such disclosure would need to be:

> *"I-8: If we want the AI to fully assist us, it's better to provide detailed information [about ourselves].*
> *I-7: So, for this news example, "DOGE plans to downsize the IRS (Internal Revenue Service) by nearly 20% ahead of tax filing season," I want to know what it means for my tax filing. I can ask AI to tell me the amount of tax I need to pay, and what consequences [will happen] if I missed anything in the current situation.*
> *I-2: Then I need to tell the AI my location, immigration status, and also my income and family situation?*
> *I-7: Yes, that helps, and it's all sensitive information.*
> *I-2: I have mixed feelings. I think personalization is good, but there are limits at some point.*
> *I-5: Of course. I understand not everyone is comfortable with that. Our information may be leaked."*

As demonstrated above, our immigrant participants were hesitant about the price of receiving personalized, action-oriented exchanges with the AI agent. Their adjusted idea, therefore, was to ensure the AI agent would run locally whenever a person sought to understand self-referential connections to the news content. This way, immigrant readers still receive actionable guidance without exposing sensitive information to a public server.

When this idea was communicated to journalists, it was refined further. A recurring question from journalists was, *"Are you sure the agent's responses are really the guidance you can follow?"* Different from explaining the composition of a dataset, which is largely fact-based and immediately verifiable, connecting a news report with an immigrant reader's life actions can often be *"interpretive rather than conclusive."* Journalists worried that AI might generate misleading responses to such inquiries, despite its ability to provide seemingly reasonable rationales.

During this part of the discussion, participants recalled that one information piece previously shared in immigrant participants' diaries was a TikToker telling his audience: *"You can skip tax filing next season because the IRS will soon be understaffed."* All participants felt alarmed at the prospect of an AI agent generating similar misleading responses. After careful consideration, immigrant readers decided to frame their envisioned AI agent as a connection informer, which underscores the agent's limited accountability in providing action suggestions to people. As I-5 elaborated:

> *"We can make this AI agent more like someone I talk to for seeing a list of possibilities, but not take them as absolute facts or guidance to follow. Instead, we think about it, collect other information, and make our own decisions. Also, it's helpful to let this agent always speak to us with a disclaimer, saying that its responses are just suggestions and it strongly encourages [the person to] double-check before acting on them." [I-5]*

***Journalist for verified implications, AI for swift initiation.*** The big group discussion with immigrant readers provoked journalists' reflections on the arguably *"vacuum space"* within their news practice: immigrant lives and perspectives are frequently missing from mainstream narratives; thus, the self-news connections outlined by an AI agent hold unique value, even though they must be interpreted with caution. All journalists emphasized that providing accountable interpretations is ultimately *"a human responsibility."* To this end, the idea of working with an AI-powered connection informer may also help improve coverage of immigrants' needs from the news production side.

Some journalists introduced a case from the COVID pandemic in the U.S., which revealed a similar *"popularity-inclusivity"* tension in their work, though under a different context. Specifically, mask wearing for health protection was a heated news topic during the pandemic. One news article that left a lasting impression on our participants asked, *"How do African Americans feel about wearing masks?"* It surfaced important insights overlooked by the majority. — For example, many African Americans reported feeling racially profiled more often because of masks; police officers might see a Black man in a mask and assume he was *"up to no good."* By describing this case, journalists explained that, even with a genuine willingness to make news content inclusive, *"the connections with someone not like myself don't always immediately click"* because people live through different experiences. Pre-discussions with an AI agent could quickly surface audience diversity to a journalist's attention. When possible, the journalist could proceed to produce a complete news report that elaborates on the information needs of a specific audience group. As J-4 speculated:

> *"If I were to use the AI agent in the newsroom, I might already have a topic focus but want different angles representing different groups of audience. I would discuss that topic with the agent, and keep digging into the suggestions it gives me until I find something newsworthy. Then, to confirm if that angle is really valid for the specific group, I must do my own research with that community. This will ensure I cover their perspectives professionally and accountably. " [J-4]*

As these discussions continued, participants from both stakeholder groups arrived at a future scenario, where a conversational AI agent acts as a **connection informer** in coordination with human stakeholders. This AI agent would run locally on an immigrant reader's device. It informs the reader, offering disclaimers, about potential connections between a given news piece and possible life actions to take. With the reader's permission, the AI agent summarizes the types of connections it had discussed with the reader and shares anonymized metadata with news outlets. By taking this latter step, the journalist would gain insights into immigrant-specific inquiries about recent news, along with any tentative implications already surfaced by the AI agent. They could then choose to investigate and produce follow-up news articles on the same topic, providing verified implications for the benefit of immigrant readers.

*4.2.3 The Empathetic Friend.* News reading triggers emotions. For immigrant readers, the emotions associated with mainstream news were largely negative, at least during the period of our research. Immigration policies underwent updates, disputes over U.S. international trade tariffs continued, and layoffs at big tech companies occurred. Participants found themselves feeling *"stressed," "unpleasant,"* and *"worried"* when following such news.

Yet they did not have the luxury of disengaging. Hard truths still conveyed information that immigrant readers needed to know. Many immigrant participants explicitly stated that they *"have to accept it even if the news just comes and slaps [them]."* Others admitted that they had thought about avoiding negative news, but soon realized they *"couldn't afford not knowing it."* Their reflections flagged the following challenge during the big group discussion: Can news be presented with more empathy so that it is less emotionally taxing to consume? Immigrant readers envisioned that a conversational AI agent could behave as an empathetic friend, easing their feelings by adjusting the presentation of given content in various ways. While journalists shared enthusiasm for this idea, they reminded immigrant participants that it would be dangerous to make serious news content overly trivialized.

***Immigrant-agent conversations to experience empathy.*** We received two top-voted ideas from immigrant participants on how they would like to experience empathy through conversations with the AI agent, thereby restoring mental energy to keep up with hard news. One idea was grounded in people's common practice of browsing a list of news at once rather than delving abruptly into one single piece. When an immigrant communicates to the AI agent that they felt anxious after browsing immigrant-unfriendly content, the AI agent could respond, *"I understand your feelings, and I want you to know that there are also positive changes around you."* In the following turns, the AI agent would guide this person toward news pieces carrying positive sentiments and discuss their meanings with the person. These purposeful combinations would enable immigrant readers to experience a more balanced set of emotions.

The other idea involves altering the presentation of the original news piece directly. Inspired by the fact that AI systems today are good at transferring the style of given content or converting texts to visuals, multiple immigrant participants imagined that the AI agent could pre-process news pieces relevant to immigrants but containing emotionally taxing content. When an immigrant communicated with the AI agent, the agent could deliver the news content in non-textual formats, for instance, *"with stories demonstrated by cute cartoon characters"* and *"with humorous narratives."* Some immigrant participants also specified that the AI-reworked format might mimic the style of satirical news shows in the U.S., such as The Daily Show or Last Week Tonight with John Oliver. As I-1 explained:

> *"For news that can deeply affect immigrants, I would prefer having technology to help me understand it in a way that does not hurt [me]. I think fun cartoons or comedy shows are good, like when there are two characters speaking. That way you can understand what happened, but you won't feel [as] unease. It won't make you enjoy what's happening [as described in the news], but it would make you feel better." [I-1]*

***Journalist for communicating facts, AI for comforting feelings.*** As much as journalists agreed with the AI agent's promise underlying both ideas, they were seriously concerned with the second one. The following reactions from J-5 and J-3, respectively, captured the common ground among all journalists in this part of the big group discussion:

> *"I like watching satirical news shows too. Constantly reading negative news in the traditional format can be disheartening, I agree. But those shows make news more palatable only after people have already engaged with the serious content. I wouldn't recommend using them to replace the serious content." [J-5]*

> *"[Presenting news content with] cartoons or characters in comics is really interesting and engaging. But to what extent does it cover the full picture? That's not enough [for news readers] to fully understand the news story. So, for me, I think people still need to click into the original news to get a full understanding… I would criticize that being too creative with the format means you are very likely to simplify the content itself. If people get used to this, they may not be able to read serious news anymore." [J-3]*

Considering views across both stakeholder groups, our participants outlined a future scenario, where a conversational AI agent acts as an **empathic friend** attending to immigrant readers' needs. In this scenario, an immigrant reader seeks help from the AI agent after encountering U.S. mainstream news that indicates negative consequences for immigrants living in the country. The AI agent would express empathy with the reader, and then comfort them by introducing other news pieces about constructive changes or favorable events. The agent might also recommend relaxing background music to accompany this immigrant's news reading; however, it would not rework any content produced by the journalist in their original news report.

*4.2.4 The Trajectory Witness.* As our participants reached the end of our big group session, the challenge at the center of their discussions gradually shifted from *"how immigrants can read each news piece more effectively"* toward *"how immigrants can become wiser news readers themselves."* We interpret this insightful shift as a consequence of the prior exchanges described under the other three metaphors. Specifically, participants began to reflect on the issue of personal agency after outlining each metaphor, noting that *"I hope we can still make sure our brains are not going to be robbed by depending on AI too much,"* and *"it feels complicated when we propose using AI to help adjust human emotions from news reading."* These reflections set the stage for their proposal of the last design metaphor: the AI agent as a trajectory witness, enabling immigrant readers' awareness, review, and analysis of their own news consumption over time. Although the remaining time in the session was not sufficient for extensive exchanges on this point, participants spent it well by generating the following thoughts.

***Immigrant-agent conversations to reflect on past trajectories.*** Conversations with the AI agent, as envisioned by immigrant readers under the other three metaphors, contain valuable and comprehensive information about their own news reading patterns and habits. These past records were viewed as these individuals' best tool to learn about themselves and, more importantly, for planning future steps toward *"[being] more sophisticated."* Immigrant participants mentioned several ways in which they would want the AI agent to analyze their news reading records, including by topical categories, by perspectives (e.g., left-wing, right-wing, or mixed), by presentation styles, and by types of questions they had raised. Several named additional analytic angles they considered important but discussed among others, such as whether the person had followed a series of reports on the same topic, or whether they had been drawn to news reports written by specific journalists.

***Journalist for guiding principles, AI for assistive practices.*** Building on the above thoughts, journalists saw potential in using such an AI agent to cultivate immigrants' literacy in consuming mainstream news. The central idea, albeit lacking in detail, was to *"add a bit more structure"* to reader-agent interactions by incorporating guidance rooted in journalists' professional knowledge. As J-1 emphasized:

> *"It would be cool if the agent can help immigrants understand the process of how news is produced, like what journalists value during the process, and why we may highlight one thing [in the news content] but not another. Then, if they want to, they can adopt similar logic [when reviewing their own news behaviors], and it can build [their] trust in news, especially in today's world with so much misinformation. I think this would require journalists to work with people who build this AI." [J-1]*

The aforementioned exchanges point to a future scenario in which a conversational AI agent serves immigrant readers as their **trajectory witness**. In this arrangement, the AI agent can walk an immigrant reader through their news reading records by drawing on past conversations with them. It assists in analytical retrospection, allowing this person to explore their news reading patterns along dimensions guided by professional principles as well as in aspects nominated by themselves. Through these practices, immigrant readers would be better equipped to consume mainstream news in their country of residence with enhanced literacy and personal agency.

***Together***, the four design metaphors portray distinct ways in which technology — particularly conversational AI agents — may contribute to enhancing reader-oriented experiences for immigrants. All of these metaphors underscore a coordinating relationship between journalists and AI agents in jointly addressing the immigrant readers' needs; meanwhile, they also demonstrate important differences in how the roles served by each party are connected.

## 5 Discussion

We reflect on the current work by revisiting our overarching inquiry: what positions should immigrant readers, journalists, and technology adopt with respect to one another in creating reader-oriented news experiences? These reflections comprise four successive lines of thought. First, we revisit the notion of immigrants as marginalized news readers, using our findings to contextualize the unique position they occupy within the mainstream news ecosystem (Section 5.1). Then, we delve into issues that would inevitably arise if either journalists or immigrant-favored technology were positioned as an isolated driver of the news experience sought by immigrants (Section 5.2). These reflections prepare us to re-examine the four participant-proposed design metaphors with a more critical lens. We consider the human-AI workflows underpinning these metaphors highly informative, particularly because each workflow illustrates a distinct and reconciled path for connecting our two stakeholders whose inherent values do not always align; nevertheless, we also note conditions that may limit a workflow from fulfilling its intended potential (Section 5.3). Several roadblocks will need to be cleared before conversational AI agents can be implemented in ways that deliver the benefits our participants envisioned. We conclude the Discussion with four concrete action items toward this end (Section 5.4).

### 5.1 News Experiences of Immigrants as a Marginalized Readership

Mainstream news often does not benefit its immigrant readers and the remaining local public equitably. Our research findings, especially those value misalignments detailed in Section 4.1, illustrate the disadvantages uniquely imposed on immigrants – *not* by considering what immigrants might gain if they were provided with additional assistance (e.g., would adding richer context improve immigrants' news reading? would it benefit local readers too?), *but* by surfacing the enduring burdens already produced by a news ecosystem that lacks so much of what is essential to immigrants (e.g., has the absence of richer context hindered immigrants' news reading? has it affected local readers to the same degree?). The former approach alone risks flattening the structural disadvantage immigrants experience, leading to a (mis)conclusion that immigrant readers need nothing really different from other readers.

Values prioritized by our immigrant participants highlight a key facet of their structural disadvantage. They inhabit a position where multiple constraints compound – being a social-status minority, navigating an often non-native language environment, *and* striving to piece together past and present societal contexts that are typically taken for granted by the local public. When local residents face difficulties in news reading, such as needing richer background information to understand a newly emerging societal issue, they can usually draw on their language proficiency and prior knowledge of the society's history to fill in the gaps. Because of the intersecting constraints they shoulder, immigrants have far fewer resources available to support comparable forms of compensation.

In addition, values prioritized by our journalist participants illuminate another essential dimension of the structural disadvantage immigrants experience. While the mainstream news system allocates resources to equip journalists with professional skills and ethics training for high-quality news production, such preparation is not designed to address the compounded constraints immigrants face on the news-consumption side. A consequence of this mismatch, evident in both our findings and prior research [36, 44, 51], is that immigrants often turn to alternative channels (e.g., explanatory YouTube videos, GPT-generated news summaries) to keep up with societal updates. Although these non-mainstream channels each respond to immigrants' needs in their own way, they lack systematic safeguards [13, 22, 91, 106]. News produced by professional journalists still carries essential qualities that cannot yet be fully replaced by such alternatives.

## 5.2 The "Unaddressed-or-Unaccountable" Paradox

To mitigate the structural disadvantage discussed above, something within today's news system must be updated to respond to immigrants' requests or to the difficulties they encounter in news reading. Both stakeholder groups in our research agree on this point; however, our findings reveal two distinct perspectives on what constitutes the first-order issue that needs to be addressed.

Our immigrant participants, throughout the entire co-design process, consistently signaled that their needs and demands as news readers remain unaddressed. In the first half of the research activities, they explicitly emphasized the anxieties caused by not being able to comprehend various aspects of meanings embedded in mainstream news. As researchers working with these participants, we interpret such anxieties as an inherent driver of the "conversational AI enthusiasm" they expressed in the second half of the process. Conversational AI agents hold the promise of always responding to immigrant readers whenever they seek information and doing so in formats adaptive to their situated requests. This responsiveness speaks directly to immigrants' concerns about repeatedly being left unaddressed within the news information loop.

A different underlying message emerged from conversations involving journalist participants: unaccountable responses are no better than no responses. The training journalists receive and uphold requires them to consistently exercise professional gatekeeping in determining what information to cover, how to cover it, and at what level the information should be processed by its readership. This strong commitment to gatekeeping matters because news, if produced without such discipline, may mislead its readers in irresponsible ways. It helps explain why journalists viewed human oversight as crucial for several aspects of news writing, such as ensuring that sensational expressions do not become manipulative, or carefully weighing information to be included or omitted in a news piece.

The significance of this "unaddressed-or-unaccountable" paradox lies in that (a) both perspectives are valid in their own contexts, and (b) neither perspective alone, in our view, can fully represent the appropriate or legitimate form of support for immigrant readers. Efforts to enhance reader-oriented experiences should not aim at simply "picking one side" or "forcing different stakeholders to reach consensus on everything." Instead, we seek to identify ways for stakeholders to collectively manage their disagreements and, in the spirit of HCI, to recognize opportunities where technology can be positioned to support this cross-stakeholder dialogue.

## 5.3 Human-AI Workflows That Underpin the Four Metaphors

The four design metaphors reported in Section 4.2 offer informative clues for managing the "unaddressed-or-unaccountable" paradox. Across these metaphors, human and AI always coordinate within workflows where both operate under boundaries carefully defined for the benefit of immigrant readers. Journalists remain responsible for ensuring the quality of foundational news content, but refraining from presuming immigrants' needs without conducting direct inquiry with this readership group. AI, in turn, conditionally augments or adapts the communication of that foundational content for end readers, and it operates with explicit disclaimers or uncertainty acknowledgments when delivering information that is interpretative in nature.

To the best of our knowledge, the current work is among the relatively small number of HCI projects where multiple stakeholder groups directly work together, transparently communicate their (mis)alignments, and collectively outline possible workflows to manage tensions. A more commonly reported practice is to engage one stakeholder group at a time, with the research team later deciding how to prioritize among different stakeholders' views and preferences (e.g., [23, 38, 47]). Although the design outcomes generated from these two processes may appear similar, we argue that the core value of research through (co-)design is to return voices to the stakeholders themselves. The human-AI workflows identified in our research carry significance largely because they emerged from constructive exchanges and negotiated inputs across both journalists and immigrant readers.

Feasibility-wise, prior work already offers a solid technical basis for many interactive features that our immigrant participants desire. For example, systems like DataDive [52] use data visualizations to help users grasp the reasoning behind a given piece of information; recommendation algorithms, once initialized with users' own inputs, show strong potential for connecting people with content aligned with their informational or affective needs [54, 60]; and recent advances in personal informatics propose frameworks for self-tracking that support reflection on past behavior and planning for long-term change [84, 97]. These forms of technical readiness can be leveraged to address immigrants' needs as articulated through the four participant-proposed metaphors.

That said, the central value of the four design metaphors lies less in the specific ways immigrants might interact with AI-powered news interfaces. Rather, it rests in the orientation toward shared accountability among journalists, immigrants, and AI systems that the workflow beneath each metaphor articulates. We note two observations with our journalist and immigrant participants that are essential for critically examining these workflows:

***Over- vs. under-estimation of AI agents' intelligence.*** Participants in both stakeholder groups lacked systematic knowledge about the limits of an AI agent's capabilities. As a result, they relied on heuristics to form expectations. Journalists, guided by their professional training and routines, tended to adopt a cautious stance toward AI, sometimes expressing concerns about accountability that felt overly conservative to immigrant participants. Immigrants, by contrast, showed greater enthusiasm for AI's potential, likely shaped by positive reinforcement from using AI tools in other work and life contexts. However, evaluation criteria from one context do not always transfer cleanly to another. For instance, while immigrant participants expressed strong confidence in AI's ability to perform style transformations on existing text, journalists emphasized that in the news domain, even small losses of nuance could introduce high-stakes risks for readers. The two groups simply did not share common reference points for assessing what AI can or should do, nor where the red lines ought to be drawn.

***Single- vs. multi-layered news information processing.*** Perhaps due to time constraints of the research sessions, participants did not discuss much about the information complexity introduced by their own design envisions. However, many of the human-AI workflows they envisioned require journalists and/or AI to layer additional information on top of the original news piece. For example, the data decoder metaphor presumes that journalists can embed multiple tiers of data explanation into the foundational news content, enabling an AI agent to surface specific layers upon the reader's request. Likewise, the connection informer and empathetic friend metaphors entail AI agents directing immigrants to supplementary information beyond the initial article. During co-design, journalists warned that AI-assisted interpretations might risk diminishing the legitimacy or flattening the nuance of the original reporting. Although immigrants agreed with this concern in group discussions, their news diary reflections prior to the sessions revealed the difficulty of digesting complex information all at once. It remains unclear how the increased information can best be displayed, organized, or reassembled for immigrant readers.

Together, these reflections suggest that our co-design findings are highly informative for guiding follow-up technology design endeavors in the news space. Nevertheless, it does not mean the four metaphors and the human-AI workflows underpinning them should be implemented as-is, without attention to their limitations.

## 5.4 Are Conversational AI Agents the Way Out?

By identifying value (mis)alignments and collectively envisioned human-AI workflows across the two stakeholder groups, our research shows that tensions are inherent to design efforts involving heterogeneous stakeholders. Rather than acting as obstacles, these tensions create opportunities to move the news ecosystem toward greater inclusion and adaptiveness for its diverse readerships. The design metaphors surfaced through our study point to concrete directions for future work, helping ensure that conversational AI agents can deliver meaningful benefits as participants envisioned:

***Equip journalists to produce AI-ready news content.*** Our findings make clear that many immigrants would like to consult conversational AI agents for urgent help with news reading even though such actions might introduce risks. Thus, we argue that the safeguarding effort for accountable news interpretation must occur upstream, at the level of journalistic practice. Journalist training in the near future should consider how foundational content may later be processed by AI agents that attempt to elaborate, rephrase, or contextualize the source content. News writing practices could therefore be upgraded to incorporate structural or metadata cues that differentiate "facts that must stay fixed" from "context that can be safely elaborated." Such practices do not change the core of journalistic work; rather, they prepare news content for downstream AI interactions that are already common in immigrants' news routine.

***Scaffold immigrants news reading with guided AI-safety protocols.*** Our immigrant participants portrayed AI agents as a reliable companion: a news reading partner they consult for summaries when short on time, explanations when confused, and reassurance when emotionally strained. While this practice speaks to the "unaddressed" side of the unaddressed-unaccountable paradox, it also heightens risks of over-reliance and misinterpretation. Follow-up design efforts should consider building accessible safety reminders into future news platforms, starting with concrete cases surfaced in our findings; for example, AI may be applied to explore how a numerical index compares historically, but not to infer the legal consequences of a civic action. Toolkits of this type would allow immigrants to continue benefiting from AI support as they already are, while maintaining an appropriate distance from outputs that may be overreaching or misleading.

***Connect journalists and immigrants for periodic reviews of AI edge cases.*** Our co-design activities revealed essential benefits of having journalists and immigrant readers examine concrete examples together, especially for cases where AI assistance can become ambiguous or risky. Whether AI agents speculated on journalists' data choices, oversimplified sensitive topics, or suggested irresponsible life decisions, these moments highlighted shared values between news content producers and consumers underneath apparent disagreements. We therefore consider a structured review of such "edge cases" as a practical step that should be built into the routine of reader-oriented news technology design. Journalists and immigrants can jointly annotate AI outputs, discuss where interpretations go astray, and identify misunderstandings or confusions. This process extends the collaborative spirit underpinning the four metaphors and generates grounded input that AI engineers can use when designing safety mechanisms.

***Document emerging improvisations of AI usage for transparency and caution alerts.*** The capabilities of AI agents continue to evolve, and so do the people who interact with them. This ongoing co-evolution means that new improvisations in AI use will inevitably emerge, and not all of them will be free of concern. Our co-design data revealed one vivid case within the news domain: journalists envisioned using AI as a shortcut to understand variation in information needs across readership groups. Similar ideas have also been explored in recent HCI work that leverages LLM-powered system infrastructures to simulate human behaviors and report notable results, primarily in terms of the human-likeness of simulated behaviors [7, 70, 71]. We argue that such improvisations must be carefully documented and evaluated with caution. As recent critical reviews emphasize (e.g., [49, 59, 96]), LLM-based AI models generate distributions of token sequences derived from training data; it does not possess human cognitive processes or the sensory and experiential grounding that shape real-world judgment. Consequently, these systems can produce highly human-like narratives without being capable of substituting for, or reliably predicting, genuine human cognition in contextualized scenarios. We recognize that both journalists and readers can benefit from receiving well-crafted, human-like text, which is already evident in the widespread use of AI-assisted editing and rephrasing tools. Yet, we remain concerned that everyday users may overextend such capabilities, introducing high-stakes risks such as diminishing the actual news experiences of marginalized readership groups, or allowing AI-generated information to overshadow journalists' own exploration of alternative framings. These concerns underscore the need to systematically document emerging improvisations in AI use, increase their transparency, and incorporate explicit cautionary alerts as new issues are identified on the fly.

We propose the above actions not only to guide the design of conversational AI agents, but also to support the ongoing coordination and calibration of the roles assigned to immigrant readers, journalists, and the technologies available to them.

## 6 Limitations and Conclusion

Similar to previous work in the news space, our research has its limitations. In particular, all insights drawn from our co-design activities were bounded by the local media landscape familiar to specific participants, as well as by the technologies feasible within their speculations. While our activity setup was effective for surfacing and connecting the perspectives of multiple stakeholders, it captured their reflections within a compressed timeframe and thus cannot fully account for the longer-term dynamics of everyday news practices. Despite these limitations, findings from this research provide a concrete account of common ground and value conflicts between news consumers and producers, distinct metaphors for news technology design, and nuanced formats of human–AI coordination that future research can further extend. By foregrounding appropriate role alignments among immigrant readers, journalists, and technology, our work underscores opportunities for more inclusive and accountable news ecosystems with AI in the loop.

## Acknowledgments

We thank Yimin Xiao, Thara Le, and Xirui Ding for their assistance with the co-design sessions. We also thank our participants for making this work possible and the anonymous reviewers for their valuable feedback. This research was supported by the National Science Foundation under grant #2443387 and #2229885. Any opinions, findings, and conclusions or recommendations expressed in this material are those of the author(s) and do not necessarily reflect the views of the National Science Foundation.